\documentclass[12pt]{article}

\usepackage{latexsym}
\usepackage{amsmath,amssymb}

\newcommand{\be}{\begin{equation}}
\newcommand{\ee}{\end{equation}}

\def\cN{{\cal N}}

\def\tr{{\rm tr}\,}
\def\Tr{{\rm Tr}\,}

\def\ui{{\underline{i}}}

\newcommand{\bea}{\begin{eqnarray}}
\newcommand{\eea}{\end{eqnarray}}
\newcommand{\ba}{\begin{array}}
\newcommand{\ea}{\end{array}}
\newcommand{\nn}{\nonumber}

\begin{document}

\begin{titlepage}

\hfill {\it Dedicated to the 60 year Jubilee of Professor D.I.
Kazakov}

\vspace{1cm}

\begin{center}
{\Large\bf Background field formalism and \\
construction of effective action for  $\cN=2$, $d=3$ \\[2mm] supersymmetric
gauge theories} \vspace{1.5cm}
 \\[0.5cm]
 {\bf
 I.L. Buchbinder $^+$\footnote{joseph@tspu.edu.ru},
 N.G. Pletnev $^\dag$\footnote{pletnev@math.nsc.ru},
 I.B. Samsonov $^{*}$\footnote{samsonov@mph.phtd.tpu.ru, on leave from
Tomsk Polytechnic University, 634050 Tomsk, Russia.}}
\\[3mm]
 {\it $^+$
 Department of Theoretical Physics, Tomsk State Pedagogical
 University,\\ Tomsk 634061, Russia \\[2mm]
 $^\dag$ Department of Theoretical Physics, Institute of Mathematics,
630090 Novosibirsk, Russia\\[2mm]
 $^*$ INFN, Sezione di Padova, via F. Marzolo 8, 35131 Padova, Italy}
 \\[0.8cm]
\bf Abstract
\end{center}
We review the background field method for three-dimensional
Yang-Mills and Chern-Simons models in $\cN=2$ superspace. Superfield
proper time (heat kernel) techniques are developed and exact
expressions of heat kernels for constant backgrounds are presented.
The background field method and heat kernel techniques are applied
for evaluating the low-energy effective actions in $\cN=2$
supersymmetric Yang-Mills and Chern-Simons models as well as in
$\cN=4$ and $\cN=8$ SYM theories.
\end{titlepage}
\setcounter{footnote}{0}

\numberwithin{equation}{section}

\section{Introduction}

It is our great pleasure to write the paper in honor of Professor
D.I. Kazakov, the bright scientist and lecturer. Professor Kazakov
was one of the pioneers in applications of superfield methods in
supersymmetric quantum field theory
\cite{Kazakov-1,Kazakov-2,Kazakov-3} and made important
contributions in the study of renormalizability properties of
three-dimensional Chern-Simons-matter theories
\cite{Kazakov1,Kazakov2}. At present, the superfield methods are
commonly recognized to be very effective for exploring quantum
aspects of supersymmetric field theories while the three-dimensional
gauge theories have again become very hot topics recently.
The present paper is devoted to a review of current state of the
problem of low-energy effective action in three-dimensional extended
supersymmetric gauge theories in the framework of $\cN=2$, $d=3$
superspace.

During last few years quantum aspects of three-dimensional
supersymmetric theories have been attracting considerable attention,
mainly because of the progress in studying field theories modelling
multiple M2 branes. The most important examples of such models are
the Bagger-Lambert-Gustavsson (BLG)
\cite{BLG,BLG1,BLG2,BLG3,BLG4,BLG5} and
Aharony-Bergman-Jafferis-Maldacena (ABJM) \cite{ABJM} theories which
are AdS/CFT-dual to the IIA superstring on the AdS$_4\times$CP$^3$
background.

On the field theory side of the AdS/CFT correspondence, major
attention is paid to the correlation functions of gauge invariant
operators. In particular, in the ABJM model such correlation
functions were studied in details \cite{Benna,MZ,GGY,GHO,GV,GM,BR,MSZ,BGR,MOSS,MOSS1,LMMOSSSTT}
(see \cite{Klose} for a
review). Another important object containing much information about quantum aspects of a field
model is the low-energy effective action. For
instance, the low-energy effective action of $\cN=4$, $d=4$ SYM
model is perfectly matched with the effective action of a probe D3 brane
moving on the AdS$_5\times$S$^5$ background \cite{CT,M,BI-review,BKT,KM,KM1,K}.
It would be very interesting to observe similar matching between
the effective action of an M2 brane on the AdS$_4\times$S$^7$
background and the low-energy effective actions of ABJM-like models.
Unfortunately, our current understanding of the latter is very
insignificant in comparison with the four-dimensional case. To
fill this gap we need to develop the methods of quantum field
theory for studying low-energy effective actions for various
three-dimensional gauge theories.

An important feature of three-dimensional gauge field theory in
comparison with the four-dimensional case is the possibility of
having topological gauge-invariant mass of gauge fields which
originates from the Chern-Simons term \cite{Deser1,Deser2,Deser3}. The Chern-Simons action,
being considered by it own, does not describe propagating degrees
of freedom and its quantization may be useful rather for
classifying topological invariants (see, e.g., \cite{Witten}). However, in models
with matter fields such Chern-Simons gauge fields are responsible for
interactions which respect conformal invariance. The ABJM and BLG
models fall exactly in this category as they represent specific
examples of Chern-Simons field theories interacting with matter in
a specific way such that the supersymmetry and conformal
invariance get enhanced.

Let us comment on the renormalizability properties of
three-dimensional gauge models. When the Chern-Simons term is
present, one should care about renormalization of the Chern-Simons level
$k$. Indeed, in the classical theory only integer values of $k$ are
compatible with large gauge invariance. In \cite{PR,CSW,CSW1,Martin1,Martin2,Kapustin-Pronin} it was proved
that the Chern-Simons level may receive only integer shifts due to
quantum corrections, so the gauge invariance is maintained at the
quantum level. This result was confirmed by many subsequent
computations for various three-dimensional gauge models with
Chern-Simons term, both with and without supersymmetry (see, e.g.,
\cite{Dunne} for a review).

More generally, one can raise the issue of finding UV finite
three-dimensional Chern-Simons-matter models which might be as
interesting as the famous $\cN=4$, $d=4$ SYM model. This problem was
extensively studied in the papers of Avdeev, Grigoriev, Kazakov and
Kondrachuk \cite{Kazakov1,Kazakov2} where RG-flows for various
three-dimensional gauge models were studied. In particular,
cancellations of two-loop divergences for $\cN=2$
Chern-Simons-matter models were found. For similar $\cN=3$
supersymmetric models a non-renormalization theorem was formulated
which proves the all-loop  UV-finiteness of such theories
\cite{BILPSZ}. These results show that non only ABJM and BLG models
should remain superconformal on the quantum level, but the class of
quantum-superconformal three-dimensional field theories is much
wider \cite{Penati1,Penati2,Penati3}. Quantum aspects and, in
particular, the problem of low-energy effective action in such
three-dimensional superconformal theories deserves much attention.

As is well known, quantization procedure of gauge theories requires
imposing a gauge which explicitly breaks the invariance of the
effective action under classical gauge transformations. To keep
track of the gauge invariance one is to employ the background field
method which was originally introduced by DeWitt \cite{DeWitt} and
developed in many subsequent papers
\cite{Kallosh1,Kallosh2,Kallosh3,Kallosh4}. The central idea of the
background field method is a decomposition of the gauge fields into
classical background and quantum fields (background-quantum
splitting) and imposing the gauge conditions only on the quantum
ones. After integrating out quantum fields, the path integral
results in the gauge invariant effective action depending on the
background fields \footnote{We stress that such an effective
action, being gauge invariant, nevertheless depends on the choice of
gauge fixing conditions (see the discussion of these aspects in
\cite{Vilkovisky}).}. However, the background-quantum splitting can
be very non-trivial for some gauge theories and, hence, the
formulation of the background field method in any concrete theory
demands a special study. For instance, construction of the
background field method for ${\cal N}=1$, $d=4$ superfield gauge
theories \cite{Grisaru} is very specific whereas its analog in the
${\cal N}=2$, $d=4$ harmonic superspace \cite{HarmBack} looks rather
similar to the one for conventional Yang-Mills theories.

The low-energy effective action usually describes an effective
dynamics of light degrees of freedom with the heavy ones
integrated out. In SYM-like gauge field theories such a
separation appears usually as a result of the Higgs mechanism of
spontaneous gauge symmetry breaking. Unfortunately, for the
ABJM-like models the Higgs mechanism works differently: once the
matter fields acquire vacua, the Chern-Simons-matter model turns
into a SYM model with higher derivative corrections \cite{M2D2,M2D2-1,M2D2-2,%
M2D2-3,M2D2-4,M2D2-5,M2D2-6,BILPSZ-ABJM}. Therefore we are led to study the
low-energy effective action in the three-dimensional SYM models
rather than in the ABJM model itself. To address the issue of effective
action in the ABJM model we need to quantize the gauge fields
in the conformal phase when all fields remain massless. Then, to
avoid IR divergences, a massive regulator (cut-off) is required.

In the present paper, after a short review of details of gauge theory in
the $\cN=2$, $d=3$ superspace given in Sect.\ 2, we develop the background field
method for $\cN=2$ super Yang-Mills and Chern-Simons models.
For these models the quadratic-fluctuations
operators are constructed for a general background and the
structure of one-loop effective action is discussed (Sect.\ 3).
Then, in Sect.\ 4 we
compute the low-energy effective actions for pure SYM models with
$\cN=2$, $\cN=4$ and $\cN=8$ supersymmetry in the Coulomb branch. For pure $\cN=2$
Chern-Simons model we compute the leading terms in the effective
action with lowest number of non-Abelian gauge superfields. For
the latter model we note that the ghost superfields at one loop produce
the $\cN=2$ SYM-like term in the effective action. This is not
surprising since the pure topological nature of classical
Chern-Simons theory is broken explicitly at the quantum level.
We believe that the presented $\cN=2$ superfield
techniques for studying effective actions in the three-dimensional
gauge models in the $\cN=2$, $d=3$ superspace will be useful for
further studies of other field theories modelling dynamics of M2
and D2 branes as well as appearing in phenomenological
applications. Some open problems which deserve further studies are
discussed in Conclusions. We follow the notations employed in
our previous works \cite{BPS1,BPS2,BP}.

\section{Supersymmetric gauge models}
\subsection{$\cN=2$, $d=3$ superspace}
The $\cN=2$, $d=3$ superspace is parametrized by the coordinates
$z^M=(x^m,\theta_\alpha,\bar\theta_\alpha)$ with
$\bar\theta_\alpha=(\theta_\alpha)^*$. The supercovariant spinor
derivatives read
\be
D_\alpha=\frac\partial{\partial\theta^\alpha}+i\bar\theta^\beta
\partial_{\alpha\beta},\qquad
\bar D_\alpha=-\frac\partial{\partial\bar\theta^\alpha}
-i\theta^\beta \partial_{\alpha\beta}\,,\qquad
\{D_\alpha, \bar D_\beta
\}=-2i\partial_{\alpha\beta}\,.
\label{Dalg}
\ee
We use the following conventions for converting the vector and bi-spinor indices to each other,
\be
x^{\alpha\beta}=(\gamma_m)^{\alpha\beta} x^m\,,\qquad
\partial_{\alpha\beta}=(\gamma^m)_{\alpha\beta}\partial_m\,,
\ee
where $(\gamma^0)_\alpha^\beta=-i\sigma_2$,
$(\gamma^1)_\alpha^\beta=\sigma_3$,
$(\gamma^2)_\alpha^\beta=\sigma_1$
are three-dimensional gamma-matrices obeying standard orthogonality and completeness relations
\be
(\gamma^m)_{\alpha\beta}(\gamma^n)^{\alpha\beta}=2\eta^{mn}\,,\qquad
(\gamma^m)_{\alpha\beta}(\gamma_m)^{\rho\sigma}
=(\delta_\alpha^\rho\delta_\beta^\sigma+\delta_\alpha^\sigma\delta_\beta^\rho)\,.
\ee
These matrices are real and symmetric. We use ``mostly minus'' Minkowski space metric,
$\eta_{mn}={\rm diag}(1,-1,-1)$.

The integration measure in the full $\cN=2$, $d=3$ superspace is
defined as
\be
d^7z\equiv d^3x d^4\theta=\frac1{16}d^3x\,D^2\bar D^2\,,\quad
\mbox{so that}\quad
\int d^3x\, f(x)=\int d^7z\,\theta^2\bar\theta^2 f(x)\,,
\label{fullmeasure}
\ee
for some field $f(x)$. Here we use the following conventions for
contractions of spinor indices
\be
D^2= D^\alpha D_\alpha\,,\quad
\bar D^2=\bar D^\alpha\bar D_\alpha\,,\quad
\theta^2=\theta^\alpha\theta_\alpha\,,\quad
\bar\theta^2=\bar\theta^\alpha\bar\theta_\alpha\,,
\ee
so that
$(\theta^2)^*=-\bar\theta^2$, $(D^2)^*=-\bar D^2$.

The chiral subspace is parametrized by $z_+=(x_+^m,\theta_\alpha)$,
where $x_\pm^m=x^m\pm i\gamma^m_{\alpha\beta}\theta^\alpha\bar\theta^\beta$.
The chiral superfields are defined as usual, $\bar
D_\alpha\Phi=0$ $\Rightarrow$ $\Phi=\Phi(x_+^m,\theta_\alpha)$. The
integration measure in the chiral superspace
$d^5z\equiv d^3x d^2\theta$ is related to the
full superspace measure (\ref{fullmeasure}) as
\be
d^7z=-\frac14d^5z\,\bar D^2=-\frac14 d^5\bar z\, D^2\,.
\label{chiral-measure}
\ee

\subsection{Gauge superfields in $\cN=2$, $d=3$ superspace}
Gauge fields in the $\cN=2$, $d=3$ superspace are introduced as the
connections for superspace derivatives (\ref{Dalg}),
\be
\nabla_\alpha=D_\alpha+A_\alpha\,,\quad
\bar\nabla_\alpha=\bar D_\alpha+\bar A_\alpha\,,\quad
\nabla_m=\partial_m+A_m\,.
\ee
These gauge connections should obey the following superfield
constraints
\cite{HKLR,N2algebra,N2algebra-1,N2algebra-2,N2algebra-3,N2algebra-4}
\bea
\{\nabla_\alpha,\bar\nabla_\beta \}&=&-2i(\gamma^m)_{\alpha\beta}
\nabla_m +2i\varepsilon_{\alpha\beta}G\,,
\label{alg1}
\\
{}[\nabla_\alpha,\nabla_m]&=&-(\gamma_m)_{\alpha\beta}\bar
W^\beta\,,\qquad
[\bar\nabla_\alpha,\nabla_m]=(\gamma_m)_{\alpha\beta}
W^\beta\,,\label{alg2}\\
{}[\nabla_m,\nabla_n]&=&i{\rm F}_{mn}\,.
\label{algebra}
\eea
Here $G$, $W_\alpha$, $\bar W_\alpha$ and ${\rm F}_{mn}$ are
non-Abelian superfield strengths subject to Bianchi
identities. In particular, the superfield $G$ is Hermitian
and is covariantly linear,
\be
\nabla^\alpha\nabla_\alpha G=0\,,\qquad
\bar\nabla^\alpha\bar\nabla_\alpha G=0\,.
\label{cov-lin}
\ee
The superfields $W_\alpha$ and $\bar W_\alpha$ are covariantly
(anti)chiral,
\be
\nabla_\alpha\bar W_\beta=0\,,\qquad
\bar \nabla_\alpha W_\beta=0
\label{cov-chiral}
\ee
and satisfy `standard' Bianchi identity,
\be
\nabla^\alpha W_\alpha=\bar\nabla^\alpha\bar W_\alpha\,.
\ee
These superfield strengths are expressed in terms of $G$ as
\be
\bar W_\alpha=\nabla_\alpha G\,,\qquad
W_\alpha=\bar\nabla_\alpha G\,.
\label{W-U}
\ee

We prefer to introduce the gauge prepotential in the
so-called chiral representation in
which the connection for the Grassmann derivative $\bar D_\alpha$
vanishes,
\be
\nabla_\alpha=e^{-2V}D_\alpha e^{2V}\,,\quad
\bar\nabla_\alpha=\bar D_\alpha \,,\quad
V^\dag=V\,.
\label{40}
\ee
In this representation the superfield strengths are expressed in
terms of the prepotential $V$ as
\be
\label{42}
G=\frac i4\bar D^\alpha(e^{-2V} D_\alpha e^{2V})\,,\quad
\bar W_\alpha=\frac i4 \nabla_\alpha \bar D^\beta (e^{-2V} D_\beta e^{2V})
\,,\quad
W_\alpha =-\frac i8 \bar D^2(e^{-2V} D_\alpha e^{2V})\,.
\ee
They are covariant under the following gauge
transformations
\be
e^{2V}\to e^{i\bar\lambda}e^{2V}e^{-i\lambda}\,,
\ee
or, in the infinitesimal form,
\be
\delta V=-\frac i2 L_V(\bar\lambda+\lambda)+\frac i2 L_V
\coth(L_V)(\bar\lambda-\lambda)\,,
\ee
where $\lambda$ and $\bar\lambda$ are chiral and antichiral
superfields, respectively, and $L_V$ denotes the commutator, e.g.,
$L_V\lambda=[V,\lambda]$.

In the Wess-Zumino gauge the component field decomposition
for $V$ is given by
\be
V=\theta^\alpha\bar\theta^\beta A_{\alpha\beta}
+ i\theta^\alpha\bar\theta_\alpha\phi
+ i\theta^2\bar\theta^\alpha\bar\lambda_\alpha
- i\bar\theta^2\theta^\alpha\lambda_\alpha
+\theta^2\bar\theta^2 D\,.
\label{V}
\ee
Here $A_{\alpha\beta}$ is a gauge vector field, $\phi$ is a real
scalar, $\lambda_\alpha$ is a complex spinor and $D$ is a real
auxiliary field.

\subsection{Super Yang-Mills model}
The classical action of the $\cN=2$ SYM can be written
equivalently either in full $\cN=2$ superspace or in the chiral
subspace,
\be
S^{\cN=2}_{\rm SYM}=\frac1{g^2}\tr\int d^7z \, G^2
=-\frac1{2g^2}\tr\int d^5z \, W^\alpha W_\alpha\,, \label{S1}
\ee
where $g$ is the dimensionfull coupling constant, $[g]=1/2$. Here
the properties (\ref{cov-lin}) and (\ref{W-U}) were used, as well
as the relation (\ref{chiral-measure}) has been applied.

The $\cN=2$ supersymmetry of the action (\ref{S1}) can be extended
to $\cN=4$ by introducing the chiral superfield $\Phi$ in the
adjoint representation of the gauge group,
\be
S^{\cN=4}_{\rm SYM}=\frac1{g^2}\tr\int d^7z \, \left( G^2
-\frac12 e^{-2V}\bar\Phi e^{2V} \Phi\right)\,,
\label{N4}
\ee
This action is invariant under non-Abelian gauge transformations
\be
\Phi\to e^{i\lambda} \Phi e^{-i\lambda}\,,\quad
\bar\Phi \to e^{i\bar\lambda} \bar\Phi e^{-i\bar\lambda}\,,\quad
e^{2V}\to e^{i\bar\lambda} e^{2V} e^{-i\lambda}\,,
\ee
with $\lambda$ and $\bar\lambda$ being (anti)chiral superfield gauge
parameters and under hidden $\cN=2$ supersymmetry,
\be
e^{-2V}\delta_\epsilon e^{2V}=\theta^\alpha\epsilon_\alpha\bar\Phi_c
-\bar\theta^\alpha\bar\epsilon_\alpha\Phi_c\,,\quad
\delta_\epsilon\Phi_c=-i\epsilon^\alpha \bar \nabla_\alpha G\,,\quad
\delta_\epsilon\bar\Phi_c=-i\bar\epsilon^\alpha \nabla_\alpha G\,.
\label{SUSY}
\ee
Here $\epsilon_\alpha$ is the anticommuting complex parameter and
$\Phi_c$, $\bar\Phi_c$ are covariantly (anti)chiral superfields,
\be
\bar\Phi_c=e^{-2V}\bar\Phi e^{2V}\,,\quad
\Phi_c=\Phi\,,\quad
\nabla_\alpha \bar\Phi_c=0\,,\quad
\bar\nabla_\alpha\Phi_c=0\,.
\ee

Similarly, the $\cN=8$ supersymmetric extension of (\ref{S1})
reads
\be
S^{\cN=8}_{\rm SYM}=\frac1{g^2}\tr\int d^7z \, \left(G^2
-\frac12 e^{-2V}\bar\Phi^i e^{2V} \Phi_i\right)
+\frac{1}{12g^2}\left(\tr\int d^5z \,\varepsilon^{ijk}
\Phi_i[\Phi_j,\Phi_k]+c.c.\right).
\label{N8}
\ee
Here $\Phi_i$, $i=1,2,3$, is a triplet of chiral superfields.
The transformations of hidden $\cN=6$ supersymmetry with the
complex anticommuting parameter $\epsilon_{\alpha\, i}$ are given
by
\bea
e^{-2V}\delta_\epsilon e^{2V}
&=&\theta^\alpha\epsilon_{\alpha\,i}\bar\Phi_c^i
-\bar\theta^\alpha\bar\epsilon_\alpha^i\Phi_{c\,i}\,,\nn\\
\delta_\epsilon\Phi_{c\,i}&=&-i\epsilon^\alpha_i \bar\nabla_\alpha G
+\frac14\varepsilon_{ijk}\bar\nabla^2(\bar\theta^\alpha\bar\epsilon_\alpha^j
\bar \Phi_c^k)
\,,\nn\\
\delta_\epsilon\bar\Phi_c^i&=&-i\bar\epsilon^{\alpha\,i} \nabla_\alpha G
+\frac14\varepsilon^{ijk}\nabla^2 (\theta^\alpha\epsilon_{\alpha\,j}\Phi_{c\,k})\,.
\label{6SUSY}
\eea
We use the standard notations $\bar\Phi_c^i=e^{-2V}\bar\Phi^i e^{2V}$,
$\Phi_c=\Phi$ for the covariantly (anti)chiral superfields.

\subsection{Chern-Simons model}

The non-Abelian $\cN=2$ supersymmetric Chern-Simons action was
constructed in \cite{N2algebra-2},
\be
S_{\rm CS}^{\cN=2}=\frac{ik}{8\pi}\tr\int_0^1 dt\int d^7z
\, \bar D^\alpha(e^{-2tV}D_\alpha e^{2tV})e^{-2tV}\partial_t
e^{2tV}\,.
\label{CS}
\ee
Here $t$ is an auxiliary real parameter and $k$ is an integer
(Chern-Simons level).

In the Abelian case the integration over the parameter $t$ can be
explicitly done,
\be
S_{\rm CS}^{\cN=2}=\frac{k}{2\pi}\int d^7z\,
VG
=\frac{k}{2\pi}\int d^3x(\frac12\varepsilon^{mnp}
A_m \partial_n A_p+i\lambda^\alpha\bar\lambda_\alpha-2\phi D)\,.
\ee
In the non-Abelian case the integration over $t$ can also be
performed for the variation of (\ref{CS}),
\be
\delta S^{\cN=2}_{\rm CS}=
\frac{k}{\pi}\tr \int d^7z
\, G\Delta V \,,
\ee
where
\be
\Delta V=\frac12e^{-2V}\delta e^{2V}
\ee
is the so-called gauge-covariant variation.

The action (\ref{CS}) allows for the $\cN=4$ supersymmetric
extension with a chiral superfield $\Phi$ in the adjoint
representation,
\be
S^{\cN=4}_{\rm CS}=S^{\cN=2}_{\rm CS}
-\frac{i k}{4\pi}\tr\int d^5z\, \Phi^2
-\frac{i k}{4\pi}\tr\int d^5\bar z \, \bar\Phi^2\,.
\label{CS4}
\ee
The transformations of hidden $\cN=2$ supersymmetry with complex
spinor parameter $\epsilon^\alpha$ read
\be
\Delta_\epsilon V=\epsilon^\alpha \bar\theta_\alpha \Phi_c
 - \epsilon^\alpha \theta_\alpha \bar\Phi_c\,,\qquad
\delta_\epsilon \Phi_c=-i\bar\epsilon^\alpha\bar\nabla_\alpha G\,,\qquad
\delta_\epsilon\bar\Phi_c=-i\epsilon^\alpha\nabla_\alpha G\,.
\label{hidden2}
\ee
Note that (\ref{hidden2}) coincides with (\ref{SUSY}) only for
real supersymmetry parameter,
$\bar\epsilon_\alpha=\epsilon_\alpha$. Therefore the sum of the
actions (\ref{N4}) and (\ref{CS4}) has $\cN=3$ supersymmetry
instead of $\cN=4$. This was first demonstrated in \cite{Kao1,Kao2} using
component field approach.

It is well known that the sum of Chern-Simons (\ref{CS}) and Yang-Mills
(\ref{S1}) actions describes topologically massive gauge theory,
\be
S[V]=S^{\cN=2}_{\rm CS}+S^{\cN=2}_{\rm SYM}\,.
\label{Ssum}
\ee
This can be most easily demonstrated for the corresponding Abelian
equation of motion,
\be
0=\frac{\delta S}{\delta V}=\frac i{g^2}\bar D^\alpha D_\alpha G
+\frac k\pi G\,.
\ee
This equation has the following differential consequence
\be
\frac{i\pi}{2kg^2}D^\alpha \bar D^2 D_\alpha G=
\bar D^\alpha D_\alpha G\,.
\ee
Now applying the identity
\be
\frac1{16}(D^2\bar D^2+\bar D^2 D^2-2D^\alpha \bar D^2
D_\alpha)=\square\,,
\ee
and using the linearity (\ref{cov-lin}) of the superfield strength
$G$ we conclude that it obeys the Klein-Gordon equation with
topological mass,
\be
(\square + m^2)G=0\,, \qquad m^2= \frac{k^2 g^4}{4\pi^2}\,.
\label{mass}
\ee
In contrast with the massless case this equation does not have
constant field solutions. This makes the perturbation theory
around classical solutions more complicate than in the pure SYM
theory.

\section{Background field method in $\cN=2$, $d=3$ superspace}
\label{BFM} Consider the $\cN=2$ super Yang-Mills-Chern-Simons model
with the action (\ref{Ssum}). As we pointed out in Introduction,
the background field method is based on a decomposition of initial
gauge field into background and quantum fields. The form of such a
splitting depends on the structure of gauge transformations. In the
theory under consideration it is convenient to decompose the gauge
superfield $V$ into the `background' $V$ and `quantum' $v$
superfields by the rule \be e^{2V}\to e^{2 V}e^{2gv}\,, \label{Vv}
\ee so that \be \nabla_\alpha= e^{-2gv}{\cal D}_\alpha
e^{2gv}\,,\qquad \bar \nabla_\alpha = \bar{\cal D}_\alpha\,, \ee
where \be {\cal D}_\alpha=e^{-2 V}D_\alpha e^{2 V}\,,\qquad
\bar{\cal D}_\alpha = \bar D_\alpha \label{calD} \ee are the
background gauge covariant spinor derivatives. There is a freedom in
defining the gauge transformations for the background and quantum
superfields. In particular, one can consider the so-called
`background' gauge transformations \be e^{2 V}\rightarrow
e^{i\bar\lambda}e^{2 V}e^{-i\lambda}\,,\qquad e^{2gv}\rightarrow
e^{i\tau}e^{2gv}e^{-i\tau} \label{bg-tr} \ee and the `quantum' ones,
\be e^{2 V}\rightarrow e^{2 V}\,,\qquad e^{2gv}\rightarrow
e^{i\bar\lambda}e^{2gv}e^{-i\lambda}\,. \label{q-tr} \ee Here
$\lambda$ and $\bar\lambda$ are (anti)chiral gauge parameters while
$\tau$ is real. The above decomposition of the gauge superfield $V$
into the background $V$ and quantum $v$ superfields is analogous to
the background-quantum splitting in ${\cal N}=1$, $d=4$ superfield
Yang-Mills theory \cite{Grisaru} but differs from the splitting in
conventional Yang-Mills theory \cite{DeWitt} and in ${\cal N}=2$,
$d=4$ super Yang-Mills theory in harmonic superspace formulation
\cite{HarmBack}.

To expand the action (\ref{Ssum}) in a series over quantum
superfields we compute the variational derivatives,
\bea \frac{\delta S}{\Delta V}&=&\frac i{g^2}{\cal
D}^\alpha\bar{\cal
D}_\alpha G+\frac k\pi G\,,\\
\frac{\delta^2 S}{\Delta V(z_1) \Delta V(z_2)}&=&
\left(\frac1{4g^2}{\cal D}^\alpha \bar{\cal D}^2{\cal D}_\alpha
-\frac{2 i}{g^2}W^\alpha {\cal D}_\alpha
+\frac{ik}{2\pi}\bar{\cal D}^\alpha {\cal D}_\alpha
\right)\delta^7(z_1-z_2)\,.
\eea
This allows us to find the leading terms with respect to the
quantum gauge superfields,
\bea
S[V,v]&=&S_0[V]+S_1[V,v]+S_2[V,v]+\ldots\,,\\
S_0[V]&\equiv& S[V]\,,\\
S_1[V,v]&=&g\,\tr\int d^7z\, v \frac{\delta S}{\Delta V}
=\tr\int d^7z\, v\left( \frac i{g}{\cal D}^\alpha\bar{\cal
D}_\alpha G+\frac{g k}\pi G \right)\,,\\
S_2[V,v]&=&\frac{g^2}2\tr \int d^7z_1 d^7z_2\, v(z_1)v(z_2)
\frac{\delta^2 S}{\Delta V(z_1)\Delta V(z_2)}\nn\\
&=&\tr \int d^7z\, v
\left(\frac1{8}{\cal D}^\alpha \bar{\cal D}^2{\cal D}_\alpha
-iW^\alpha {\cal D}_\alpha
+\frac{ig^2k}{4\pi}\bar{\cal D}^\alpha {\cal D}_\alpha \right)v\,.
\eea
Here we do not consider the terms with vertices for
quantum superfields as we restrict ourself to one-loop
computations only.

The action $S_1$ is responsible for the equations of motion for
the background superfield and does not contribute to the effective
action. The action $S_2$ can be rewritten in the form
\be
S_2=\tr\int d^7z\, v(H_1+H_2)v\,,
\ee
where the operators $H_1$ and $H_2$ originate from the second
variations of the SYM and Chern-Simons actions, respectively,
\be
H_1=\frac1{4}{\cal D}^\alpha \bar{\cal D}^2{\cal D}_\alpha
-iW^\alpha {\cal D}_\alpha\,,\qquad
H_2=\frac{ig^2k}{4\pi}\bar{\cal D}^\alpha {\cal D}_\alpha\,.
\ee
Both these operators are degenerate. By fixing the gauge we can
remove the degeneracy of either of these operators. We will make
them both non-degenerate since this option is more general.

Within the background field method
one usually fixes the quantum gauge symmetry (\ref{q-tr}) keeping the
invariance under the background transformations. The corresponding
gauge fixing functions
\be
f= i\bar{\cal D}^2 v\,,\qquad
\bar f=i{\cal D}^2 v
\label{ff}
\ee
are defined with the help of the background-dependent covariant
spinor derivatives (\ref{calD}). These functions are covariantly (anti)chiral
and change under the quantum gauge transformations (\ref{q-tr}) as
\be
\delta f=\frac 1{2g}\bar {\cal D}^2 L_{gv}[\bar\lambda+\lambda
+\coth(L_{gv})(\lambda-\bar\lambda)]\,.
\ee
Therefore the ghost superfield action has the standard form,
\be
S_{\rm gh}=\tr\int d^7z\,(b+\bar b)
L_{gv}[c+\bar c+\coth (L_{gv})(c-\bar c)]
=\tr\int d^7z\,(\bar b c-b\bar c)+O(g)\,.
\ee

The one-loop effective action is given by the following functional
integral
\be
e^{i\Gamma[V]}=e^{iS[V]}\int {\cal D}v{\cal D}b
{\cal D}c\, \delta[f-i\bar{\cal D}^2 v]\delta[\bar f-i{\cal D}^2 v]
e^{iS_2[V,v]+iS_{\rm gh}}\,.
\ee
To represent the delta-functions in Gaussian form we
average this expression with the weight
\be
1=\int {\cal D}f{\cal D}\varphi\, e^{iS[f]+iS[\varphi]},
\ee
where
\be
S[f]=\frac1{8\alpha}\tr\int d^7z \, f\bar f
+\frac i{4\beta} \tr\int d^5z\, f^2+\frac i{4\beta}
 \tr \int d^5\bar z\, \bar f^2\,,
\label{Sf}
\ee
and the action $S[\varphi]$ coincides with (\ref{Sf}), but depends
on the anticommuting Nielsen-Kallosh ghost $\varphi$. Parameters $\alpha$ and
$\beta$ in (\ref{Sf}) are arbitrary. For finite values
of these parameters the action (\ref{Sf}) describes the massive Wess-Zumino
model, but one can eliminate either $f\bar f$ or massive terms by
sending corresponding parameter to infinity. As a result, we get
the gauge fixing and Nielsen-Kallosh actions in the form
\bea
S_{\rm gf}&=&\tr\int d^7z\, v\left[ -\frac1{16\alpha}\{{\cal D}^2,\bar{\cal D}^2\}
+\frac i\beta \bar {\cal D}^2 +\frac i\beta {\cal D}^2
\right]
v\,,\\
S_{\rm NK}&=&-\tr\int d^7z \,\bar\varphi\varphi\,.
\label{S-NK}
\eea
Note that the mass term for the Nielsen-Kallosh ghost vanishes
because of anticommuting nature of this superfield.

One of the most simple choice for the gauge fixing parameters
corresponds to $\alpha=1$ and $\beta=\frac{8\pi}{kg^2}$.
Then the sum of the actions $S_2$ and $S_{\rm gf}$ reads
\be
S_2+S_{\rm gf}=\tr\int d^7z \, v(
-\square_{\rm v}+H)v\,,
\ee
where
\bea
\square_{\rm v}&=&-\frac18 {\cal D}^\alpha \bar{\cal D}^2{\cal D}_\alpha
+\frac1{16}\{{\cal D}^2,\bar{\cal D}^2  \}
+\frac i2({\cal D}^\alpha{W}_\alpha)+i{W}^\alpha{\cal
D}_\alpha\nn\\
&=&{\cal D}^m {\cal D}_m +{G}^2 +i{W}^\alpha{\cal D}_\alpha
-i\bar{W}^\alpha\bar{\cal D}_\alpha\,,
\label{v-square}
\\
H&=&\frac{ig^2 k}{8\pi}(
\bar {\cal D}^\alpha {\cal D}_\alpha
+ {\cal D}^\alpha \bar {\cal D}_\alpha+
\bar{\cal D}^2 + {\cal D}^2)\,.
\label{H}
\eea
Here $\square_{\rm v}$ is the covariant d'Alembertian operator in the space of real
superfields and $H$ originates from the Chern-Simons part of the
action.

As a result, we get the following representation for the one-loop
effective action
\be
e^{i\Gamma[{V}]}=e^{iS[{V}]}\int {\cal D}v{\cal D}b
{\cal D}c{\cal D}\varphi
e^{i\tr\int d^7z\, v(-\square_{\rm v}+H)v+iS_{\rm gh}+iS_{\rm NK}}\,.
\label{75}
\ee
Schematically, it can be written as
\be
\Gamma=\Gamma_{\rm v}+\Gamma_{\rm gh}\,,\qquad
\Gamma_{\rm v}=\frac i2{\rm Tr_v}\ln( \square_{\rm v}-H)\,,\quad
\Gamma_{\rm gh}=-\frac{3i}2{\rm Tr}_+\ln \square_+\,.
\label{2.31}
\ee
The contribution $\Gamma_{\rm v}$ to the one-loop effective action
comes from the quantum gauge superfield while $\Gamma_{\rm gh}$ is due to ghosts.
Here ${\rm Tr_v}$
and ${\rm Tr}_+$ are the functional traces of the operators acting in
the spaces of real and chiral superfields, respectively.
The operator $\square_+$ is the covariant d'Alembertian operator
acting in the space of covariantly chiral superfields
which was introduced in \cite{BPS2},
\be
\square_+=\frac1{16}\bar{\cal D}^2{\cal D}^2={\cal D}^m{\cal D}_m+G^2+\frac i2({\cal D}^\alpha {W}_\alpha)
+i{W}^\alpha {\cal D}_\alpha\,.
\label{square+}
\ee
The explicit expressions for the traces of these operators can be
found after one specifies the gauge group and the
background gauge superfield.

\section{Superfield effective action}
\subsection{$\cN=2$ SYM}
\label{sect2.3}
Consider pure $\cN=2$ SYM model with classical action (\ref{S1}).
The background field method goes along the same lines as in Sect.\
\ref{BFM}, but in eq.\ (\ref{Sf}) we send $\beta\to\infty$ to
remove the part of this action responsible for the gauge fixing in
the Chern-Simons action. Then the expression for the one-loop effective action
(\ref{2.31}) slightly modifies,
\be
\Gamma_{\rm SYM}^{\cN=2}=\Gamma_{\rm v}+\Gamma_{\rm gh}\,,\qquad
\Gamma_{\rm v}=\frac i2{\rm Tr_v}\ln \square_{\rm v}\,,\quad
\Gamma_{\rm gh}=-\frac{3i}2{\rm Tr}_+\ln \square_+\,,
\label{2.32}
\ee
where the operators $\square_{\rm v}$ and $\square_+$ are given in
(\ref{v-square}) and (\ref{square+}).

We will be interested in the low-energy effective action
which is a functional for the massless fields obtained by
integrating out all massive ones in the functional integral. In
gauge theories the separation between massless and massive
fields appears usually through the Higgs mechanism.
In general, the gauge group SU$(N)$ is spontaneously broken down to its
maximal Abelian subgroup, U$(1)^{N-1}$. However, in particular
cases a bigger subgroup of SU$(N)$ can be unbroken. Physically
interesting to consider minimal gauge symmetry breaking,
${\rm SU}(N)\to {\rm SU}(N-1)\times {\rm U}(1)$ because,
from the point of view of D-branes, the corresponding
effective action contains the potential which appears when
one separates one D-brane from the stack. In this section we will
consider first the general case when the gauge group is broken down to
the maximal torus and then comment on the effective action with minimal
gauge symmetry breaking.

The Lie algebra su$(N)$ consists of Hermitian traceless
matrices. Any element $v$ of su$(N)$ can be represented by a
decomposition over the Cartan-Weil basis in the gl$(N)$ algebra,
\be
(e_{IJ})_{LK}=\delta_{IL}\delta_{JK}\,,\qquad
v=\sum_{I<J}^N (v_{IJ} e_{IJ}+\bar v_{IJ}e_{JI})
+\sum_{I=1}^N v_I e_{II}\,,\quad \bar v_I=v_I\,,\quad \sum_{I=1}^N v_I=0\,.
\label{basis}
\ee
The background gauge superfield $V$ belongs to the Cartan
subalgebra spanned on $e_{II}$,
\be
V=\sum_{I=1}^N {\bf V}_I e_{II}={\rm diag}({\bf V}_1,{\bf V}_2,
\ldots,{\bf V}_N)\,,\quad
\bar {\bf V}_I={\bf V}_I\,,\quad \sum_{I=1}^N {\bf
V}_I=0 \,.
\label{V-back}
\ee
In what follows we will denote by boldface Latin letters the
matrix elements of the background superfields. In particular, each
matrix element ${\bf V}_I$ of $V$
has superfield strength ${\bf G}_I=\frac i2\bar D^\alpha D_\alpha {\bf
V}_I$ which is computed as in the Abelian case.
We will use also the following notations
\be
{\bf V}_{IJ}={\bf V}_I-{\bf V}_J\,,\quad
{\bf G}_{IJ}={\bf G}_I-{\bf G}_J\,,\quad
{\bf W}_{IJ\,\alpha}={\bf W}_{I\,\alpha}-{\bf W}_{J\,\alpha}\,.
\ee

Now we can do the matrix trace in the quadratic action,
\be
S_2+S_{\rm gf}=-\tr\int d^7z \, v
\square_{\rm v}v=-2\sum_{I<J}^N\int d^7z\,
v_{IJ}\hat\square_{{\rm v}\,IJ}\, \bar v_{IJ}\,,
\ee
where
$\hat\square_{{\rm v}\,IJ}$ is the Abelian version of the operator
(\ref{v-square}) which is constructed from the Abelian gauge
superfield ${\bf V}_{IJ}$ and its superfield strengths,
\be
\hat\square_{{\rm v}\,IJ} ={\cal D}^m {\cal D}_m +{\bf G}_{IJ}^2 +i{\bf
W}_{IJ}^\alpha{\cal D}_\alpha -i\bar{\bf W}_{IJ}^\alpha\bar{\cal
D}_\alpha\,.
\label{v-square-ij}
\ee
Therefore the effective action
$\Gamma_{\rm v}$ can be written as
\be
\Gamma_{\rm v}=i\sum_{I<J}^N {\rm Tr_v} \ln \hat\square_{{\rm v}\,IJ}\,,
\label{Gamma-ij}
\ee
where ${\rm Tr_v}$ means now only the functional trace in the space of real
superfields.

In a similar way one can analyze the contributions from the ghost
superfields. Consider, for instance, the action for the
Nielsen-Kallosh ghost (\ref{S-NK}) in which the chiral superfields
are expanded over the basis (\ref{basis}) as \be \varphi=\sum_{I\ne
J}^N e_{IJ} \varphi_{IJ}\,,\qquad \bar\varphi=\sum_{I\ne J}^N
e_{IJ}\bar \varphi_{IJ}\,. \label{q-varphi} \ee Here we omit the
diagonal components because they do not interact with the background
gauge superfield (\ref{V-back}) and do not contribute to the
effective action. Then the matrix trace in the action (\ref{S-NK})
is done,
\be
S_{\rm NK}=-\sum_{I\ne J}^N\int d^7z\,
\bar\varphi_{IJ}\varphi_{IJ}\,,
\ee
where the superfields
$\bar\varphi_{IJ}$ are covariantly antichiral, \be e^{-2{\bf
V}_{IJ}}D_\alpha e^{2{\bf V}_{IJ}}\bar \varphi_{IJ}=0\ \mbox{ for
}I<J\,,\qquad e^{2{\bf V}_{IJ}}D_\alpha e^{-2{\bf V}_{IJ}}\bar
\varphi_{IJ}=0\ \mbox{ for }I>J\,. \ee We see that the chiral
superfields appear in pairs with positive and negative charges with
respect to the Abelian gauge superfield ${\bf V}_{IJ}$. This
prevents the generation of the Chern-Simons term in the one-loop
computations (there is no parity anomaly \cite{anomaly,anomaly-1,anomaly-2}).
As a result, the
effective action for the ghost superfields reads
\be \Gamma_{\rm gh}=-3i\sum_{I<J}^N{\rm Tr}_+ \ln \hat\square_{+IJ}\,, \label{Gamma+ij}
\ee
where $\hat\square_{+IJ}$ is the Abelian version of the operator
(\ref{square+}) constructed from the gauge superfield ${\bf
V}_{IJ}$,
\be
\hat\square_{+IJ}={\cal D}^m{\cal D}_m+{\bf G}_{IJ}^2
+\frac i2({\cal D}^\alpha {\bf W}_{IJ\, \alpha}) +i{\bf
W}^\alpha_{IJ} {\cal D}_\alpha\,,
\label{square-ij}
\ee
and ${\rm Tr}_+$ denotes the functional trace in the space of chiral superfields.

To do the explicit quantum computations of traces of logarithms in
(\ref{Gamma-ij}) and (\ref{Gamma+ij}) we have to specify the constraints on the
background Abelian superfields:\\
(i) The matrix components of the background gauge superfield ${\bf
V}_{IJ}$ obey the $\cN=2$ supersymmetric Maxwell equations,
\be
D^\alpha{\bf W}_{IJ\,\alpha}
=\bar D^\alpha\bar {\bf W}_{IJ\,\alpha}=0\,.
\label{approx-1}
\ee
(ii) We study the effective action in the
so-called long-wave approximation in which the space-time
derivatives of the background are neglected,
\be
\partial_m{\bf G}_{IJ}=\partial_m {\bf W}_{IJ\,\alpha}
=\partial_m \bar {\bf W}_{IJ\,\alpha}=0\,.
\label{approx-2}
\ee
For such a background the heat kernels of the operators
(\ref{v-square-ij}) and (\ref{square-ij}) are known, see
\cite{BPS2}. In the present notations they read
\bea
K_{{\rm v}\,IJ}(z,z'|s)
&=&\frac1{8(i\pi s)^{3/2}}\frac{s{\bf B}_{IJ}}{\sinh(s{\bf B}_{IJ})}
e^{is{\bf G}_{IJ}^2}
e^{\frac i4({\bf F}_{IJ}\coth s{\bf F}_{IJ})_{mn}\zeta^m(s)\zeta^n(s)}
\zeta^2(s)\bar\zeta^2(s)\,,\\
K_{+IJ}(z,z'|s)&=&-\frac14\bar{\cal D}^2 K_{{\rm
v}IJ}(z,z'|s)\,,
\eea
where ${\bf B}_{IJ}^2=\frac12 D_\alpha{\bf W}_{IJ}^\beta D_\beta
{\bf W}_{IJ}^\alpha$ and
\be
\zeta^m=(x-x')^m-i\zeta \gamma^m \bar\theta'+i\theta'\gamma^m
\bar\zeta\,,\quad
\zeta^\alpha=\theta^\alpha-\theta'^\alpha\,,\quad
\bar\zeta^\alpha=\bar\theta^\alpha-\bar\theta'^\alpha
\label{interval}
\ee
are the components of supersymmetric interval.
In fact, for the one-loop computations we need these
expressions only at coincident superspace points,
\bea
K_{{\rm v}\,IJ}(s)&\equiv&K_{{\rm v}\,IJ}(z,z|s)=\frac1{(i\pi)^{3/2}}\frac1{\sqrt s}
\frac{{\bf W}_{IJ}^2\bar{\bf W}_{IJ}^2}{{\bf B}_{IJ}^3}e^{is{\bf G}_{IJ}^2}
\tanh\frac{s{\bf B}_{IJ}}{2}\sinh^2\frac{s{\bf B}_{IJ}}{2}\,,\\
K_{+IJ}(s)&\equiv&K_{+IJ}(z,z|s)=\frac1{8(i\pi s)^{3/2}}s^2{\bf W}_{IJ}^2
e^{is{\bf G}_{IJ}^2}
\frac{\tanh(s{\bf B}_{IJ}/2)}{s{\bf B}_{IJ}/2}\,.
\label{K+}
\eea
The corresponding contributions to the effective action from these heat
kernels are given by
\be
\Gamma_{\rm v}=-i\sum_{I<J}^N
\int_0^\infty\frac{ds}s\int d^7z\,K_{{\rm v}\,IJ}(s)\,,\qquad
\Gamma_{\rm gh}=-3i\sum_{I<J}^N
\int_0^\infty\frac{ds}s\int d^5z\,K_{+IJ}(s)\,,
\ee
or, explicitly,
\bea
\Gamma_{\rm v}&=&-\frac1\pi\sum_{I<J}^N
\int d^7z\int_0^\infty\frac{ds}{s\sqrt{i\pi s}}
\frac{{\bf W}_{IJ}^2\bar{\bf W}_{IJ}^2}{{\bf B}_{IJ}^3}e^{is{\bf G}_{IJ}^2}
\tanh\frac{s{\bf B}_{IJ}}{2}\sinh^2\frac{s{\bf B}_{IJ}}{2}\,,
\label{Gv}\\
\Gamma_{\rm gh}&=&-\frac{3}{2\pi}\sum_{I<J}^N
\int d^7z\bigg[{\bf G}_{IJ}\ln{\bf G}_{IJ}\nn\\&&
+\frac14\int_0^\infty\frac{ds}{\sqrt{i\pi s}}
e^{is{\bf G}_{IJ}^2}\frac{{\bf W}_{IJ}^2\bar{\bf W}_{IJ}^2}{{\bf B}_{IJ}^2}\left(
\frac{\tanh(s{\bf B}_{IJ}/2)}{s{\bf B}_{IJ}/2}-1
\right)\bigg],\label{Gh}
\eea
where in the expression for $\Gamma_{\rm gh}$ we restored the
full superspace measure. The sum of the expressions (\ref{Gv}) and (\ref{Gh})
gives us the resulting one-loop effective action in the pure
$\cN=2$ SYM theory for the gauge group SU$(N)$ spontaneously
broken down to U$(1)^{N-1}$. We point out
that only the leading ${\bf G}\ln\bf G$ term  in the $\cN=2$ SYM
effective action was obtained in \cite{deBoer,deBoer-1}
using the duality transformations while the explicit quantum
computations allow us to find all higher-order $F^{2n}$ terms
encoded in the proper-time integrals (\ref{Gv}) and (\ref{Gh}).

Let us comment on the case of minimal
gauge symmetry breaking ${\rm SU}(N)\to {\rm SU}(N-1)\times {\rm
U}(1)$. In this case it is convenient to choose the background gauge
superfield in the following form
\be
V=\frac1{N}{\rm
diag}\left((N-1){\bf V}, \underbrace{-{\bf V},\ldots,-{\bf
V}}_{N-1}\right),
\label{V-back1}
\ee
where $\bf V$ is Abelian gauge
superfield with the superfield strengths $\bf G$, ${\bf W}_\alpha$ and
$\bar{\bf W}_\alpha$. One can easily repeat all the above
considerations for such a background or just extract the answer
from (\ref{Gv}) and (\ref{Gh}) by substituting the corresponding
expressions for ${\bf V}_{IJ}$. For simplicity, we give here only
two leading terms in the corresponding effective action
\be
\Gamma^{\cN=2}_{\rm SYM}= -\frac{3(N-1)}{2\pi}\int d^7z\,{\bf
G}\ln{\bf G} +\frac{9(N-1)}{128\pi}\int d^7z\frac{{\bf
W}^2\bar{\bf W}^2}{{\bf G}^5} +\ldots\,.
\label{G-N2}
\ee
The first
term in the rhs is responsible for the $\cN=2$ supersymmetric (and
superconformal) generalization of the Maxwell $F^2$ term while the
second one gives $F^4$ among other components. The dots here stand
for higher orders of the Maxwell field strength.

\subsection{$\cN=4$ SYM}
Consider the $\cN=4$ SYM model with the classical action
(\ref{N4}). We have to extend the background field method
presented in Sect.\ \ref{BFM} with the corresponding
background-quantum splitting for the chiral superfield,
\be
\Phi\to \Phi+g\phi\,,\qquad
\bar\Phi\to \bar\Phi+g\bar\phi\,.
\label{Phi-phi}
\ee
Here the superfields $\Phi$, $\phi$ and $\bar\Phi$, $\bar\phi$
in the right hand sides are covariantly (anti)chiral
with respect to the background gauge covariant
derivatives, ${\cal D}_\alpha\bar\Phi={\cal D}_\alpha\bar\phi=0$,
$\bar{\cal D}_\alpha\Phi=\bar{\cal D}_\alpha\phi=0$.
The quantum gauge transformations for these
superfields read
\be
\delta \phi=i[\lambda,\frac1g\Phi+\phi]\,,\quad
\delta\bar\phi=i[\bar\lambda,\frac1g\bar\Phi+\bar\phi]\,,\quad
\delta\Phi=\delta\bar\Phi=0\,.
\label{q-tr1}
\ee

Upon the background-quantum splitting (\ref{Vv}) and (\ref{Phi-phi}), the
$\cN=4$ SYM action (\ref{N4}) can be expanded in a series over the quantum
superfields. In particular, for the one-loop computations we need
the quadratic part of this action,
\bea
S_2&=&
-\tr\int d^7z\,v
[-\frac18 {\cal D}^\alpha \bar{\cal D}^2{\cal D}_\alpha
+\frac i2({\cal D}^\alpha W_\alpha)+i W^\alpha{\cal
D}_\alpha+\Phi\bar\Phi]v\nn\\
&&-\tr\int d^7z(-\bar\phi[\Phi,v]+\phi[\bar\Phi,v]
+\frac12\phi\bar\phi)\,.
\label{3.7}
\eea
This action is invariant under the quantum gauge transformations
(\ref{q-tr}) and (\ref{q-tr1}). Therefore we fix the quantum gauge
symmetry by the following gauge-fixing functions
\be
f= i\bar{\cal D}^2 v-\frac i2[\Phi,\bar
{\cal D}^2\square_-^{-1}\bar\phi]\,,\qquad
\bar f=i{\cal D}^2 v+\frac i2[\bar\Phi,{\cal D}^2\square_+^{-1}\phi]
\,.
\label{f-f}
\ee
In comparison with (\ref{ff}), these functions have the terms
depending on the background (anti)chiral superfields $\Phi$ and
$\bar\Phi$ which are necessary to remove the mixed terms between
the quantum gauge $v$ and (anti)chiral $\bar\phi$, $\phi$
superfields. Such a gauge fixing is usually referred to as the
generalized $R_\xi$ gauge \cite{BBP,Ovrut,PSS}.
The corresponding gauge-fixing action reads
\bea
S_{\rm gf}&=&\frac 18\tr\int d^7z \,\bar f f
=\frac 1{8}\tr\int d^7z\bigg(-\frac12 v\{{\cal D}^2,\bar{\cal D}^2 \}v
-\frac12 v\bar{\cal D}^2[\bar\Phi,{\cal D}^2\square_+^{-1}\phi]
\nn\\&&
+\frac12 v{\cal D}^2[\Phi,\bar{\cal D}^2\square_-^{-1}\bar\phi]
+\frac14[\Phi,\bar{\cal D}^2\square_-^{-1}\bar\phi]
[\bar\Phi,{\cal D}^2\square_+^{-1}\phi]
\bigg)\,.
\label{3.9}
\eea

It is convenient at this point to specify the constraints on the
background chiral superfields $\Phi$ and $\bar\Phi$,
\be
{\cal D}_\alpha\Phi=0\,,\qquad
\bar{\cal D}_\alpha\bar\Phi=0\,,
\label{approx-3}
\ee
i.e. they are covariantly constant.
For such a background the action (\ref{3.9}) simplifies,
\be
S_{\rm gf}=\tr\int d^7z\left(-\frac1{16} v\{{\cal D}^2,\bar{\cal D}^2 \}v
- v [\bar\Phi,\phi]
+ v[\Phi,\bar\phi]
+\frac12[\Phi,\square_-^{-1}\bar\phi]
[\bar\Phi,\phi]
\right)\,.
\ee
As a result, the quadratic part of the action for the quantum
superfields becomes very simple,
\be
S_2+S_{\rm gf}=-\tr\int d^7z\left[
v(\square_{\rm v}+\bar\Phi\Phi)v+
\frac12\phi(1+\bar\Phi\Phi\square_-^{-1})\bar\phi
\right].
\ee
Here we denote $\bar\Phi\Phi v=[\bar\Phi,[\Phi, v]]$ and
$\bar\Phi\Phi \bar\phi=[\bar\Phi,[\Phi,\bar\phi]]$.

The quantum gauge transformations (\ref{q-tr}) and (\ref{q-tr1})
define the action for the ghost superfields,
\bea
S_{\rm gh}&=&\tr\int d^7z(b+\bar b)
L_{gv}[c+\bar c+\coth (L_{gv})(c-\bar c)]\nn\\&&
-\tr\int d^7z \left(
b[\Phi,\square_-^{-1}[\bar\Phi+g\bar\phi, \bar c]]
-\bar b[\bar \Phi,\square_+^{-1}[\Phi+g\phi,c]]
\right)\nn\\
&&-\tr\int d^7z\,\bar\varphi\varphi\,. \eea Here $b$ and
$c$ are standard Faddeev-Popov ghosts while $\varphi$ is the
Nielsen-Kallosh ghost. All these superfields are covariantly
(anti)chiral. Up to the second order in quantum superfields, the
ghost superfield action is given by
\be
S_{\rm gh}=\tr\int d^7z\left[ -b(1+\Phi\square_-^{-1}\bar\Phi)\bar c +\bar
b(1+\bar\Phi\square_+^{-1}\Phi)c -\bar\varphi\varphi\right].
\label{gh-4}
\ee

The functional integral for the one-loop effective action reads
\be
e^{i\Gamma_{\rm SYM}^{\cN=4}[{V},\Phi]}=e^{iS^{\cN=4}_{\rm SYM}[{V},\Phi]}
\int {\cal D}v{\cal D}\phi {\cal D}b
{\cal D}c{\cal D}\varphi
e^{iS_2+iS_{\rm gf}+iS_{\rm gh}}\,.
\ee
Schematically, the one-loop effective action can be written as
\be
\Gamma_{\rm SYM}^{\cN=4}=\frac i2{\rm Tr_v}\ln(\square_{\rm v}+\bar\Phi\Phi)
-i{\rm Tr}_+\ln (\square_++\bar\Phi\Phi)\,.
\label{N4EA}
\ee
The first term in the rhs in this expression comes from the
quantum gauge superfield while the second one takes into account
the contributions from quantum chiral superfield $\phi$ and ghosts.

Now let us compute the traces of the logarithms of the operators
in (\ref{N4EA}) for the gauge group SU$(N)$ spontaneously broken
down to U$(1)^{N-1}$. The background gauge superfield $V$ is
specified in (\ref{V-back}). The background chiral superfield $\Phi$ has
similar structure,
\be
\Phi={\rm diag}({\bf\Phi}_1,{\bf\Phi}_2,\ldots,{\bf\Phi}_N)\,,\qquad
\sum_{I=1}^N {\bf\Phi}_I=0\,.
\label{phi-back}
\ee
The quantum gauge superfield $v$ is given by the expansion (\ref{basis})
while the quantum chiral superfield $\phi$ is represented by the expression similar
to (\ref{q-varphi}). It is straightforward to compute the
matrix traces in (\ref{N4EA}),
\be
\Gamma^{\cN=4}_{\rm SYM}=i\sum_{I<J}^N{\rm Tr_v}\ln(
\hat\square_{{\rm v}\,IJ}+\bar{\bf\Phi}_{IJ}{\bf\Phi}_{IJ})
-2i\sum_{I<J}^N{\rm Tr}_+\ln (\hat\square_{+IJ}+\bar{\bf\Phi}_{IJ}
{\bf\Phi}_{IJ})\,,
\ee
where ${\bf\Phi}_{IJ}={\bf\Phi}_I-{\bf\Phi}_J$ and the operators
$\hat\square_{{\rm v}\,IJ}$ and $\hat\square_{+IJ}$ are given in
(\ref{v-square-ij}) and (\ref{square-ij}), respectively. The traces of the
logarithms of these operators are computed in a similar way as in
Sect.\ \ref{sect2.3}. As a result we get the one-loop effective
action in the $\cN=4$ SYM theory for the gauge group SU$(N)$ broken down
to U$(1)^{N-1}$,
\bea
\Gamma^{\cN=4}_{\rm SYM}
&=&-\frac1\pi\sum_{I<J}^N\int d^7z\int_0^\infty\frac{ds}{s\sqrt{i\pi s}}
\frac{{\bf W}_{IJ}^2\bar{\bf W}_{IJ}^2}{{\bf B}_{IJ}^3}
e^{is({\bf G}_{IJ}^2+\bar{\bf \Phi}_{IJ}{\bf\Phi}_{IJ})}
\tanh\frac{s{\bf B}_{IJ}}{2}\sinh^2\frac{s{\bf B}_{IJ}}{2}\nn\\&&
-\frac{2}{\pi}\sum_{I<J}^N\int d^7z\bigg[
 {\bf G}_{IJ}\ln({\bf G}_{IJ}+\sqrt{{\bf G}_{IJ}^2+\bar{\bf \Phi}_{IJ}{\bf \Phi}_{IJ}})
  - \sqrt{{\bf G}_{IJ}^2+\bar{\bf \Phi}_{IJ}{\bf \Phi}_{IJ}}\nn\\&&
+\frac14\int_0^\infty\frac{ds}{\sqrt{i\pi s}}
e^{is({\bf G}_{IJ}^2+\bar{\bf \Phi}_{IJ}{\bf \Phi}_{IJ})}\frac{{\bf W}_{IJ}^2\bar{\bf W}_{IJ}^2}{{\bf B}_{IJ}^2}\left(
\frac{\tanh(s{\bf B}_{IJ}/2)}{s{\bf B}_{IJ}/2}-1
\right)\bigg].
\label{Gamma-N4}
\eea
We point out that only the leading terms given in the second line
in (\ref{Gamma-N4}) were studied in \cite{deBoer,deBoer-1} by
employing the mirror symmetry while here we computed also all
higher order terms which are responsible in components for all higher powers
of the Maxwell field strength $F^{2n}$, $n\geq2$.

In conclusion of this section let us briefly comment on the case
of minimal gauge symmetry breaking ${\rm SU}(N)\to {\rm SU}(N-1)\times {\rm
U}(1)$. The background chiral superfield $\Phi$ is chosen
similarly as the gauge one (\ref{V-back1}),
\be
\Phi=\frac1{N}{\rm diag}\left((N-1){\bf \Phi},
\underbrace{-{\bf \Phi},\ldots,-{\bf \Phi}}_{N-1}\right).
\label{Phi-back2}
\ee
The leading terms in the $\cN=4$ SYM effective action in this case
are given by
\be
\Gamma^{\cN=4}_{\rm SYM}
=\frac{2(N-1)}{\pi}\int d^7z\bigg[
  \sqrt{{\bf G}^2+\bar{\bf \Phi}{\bf \Phi}}
  - {\bf G}\ln({\bf G}+\sqrt{{\bf G}^2+\bar{\bf \Phi}{\bf \Phi}})
+\frac{1}{32}\frac{{\bf W}^2\bar{\bf W}^2}{({\bf G}^2+\bar{\bf \Phi}{\bf\Phi})^{5/4}}
+\ldots\bigg].
\label{3.21}
\ee
The first two terms in the rhs of this expression are responsible
for $\cN=4$ supersymmetric (and superconformal) generalization of the
Maxwell $F^2$ term
while the third term gives $F^4$ among other components
and the dots stand for higher-order terms.

Finally, let us comment on the following terms in the effective action
(\ref{3.21}),
\be
\int d^7z[
{\bf G}\ln({\bf G}+\sqrt{{\bf G}^2+\bar{\bf \Phi}{\bf \Phi}})
  - \sqrt{{\bf G}^2+\bar{\bf \Phi}{\bf \Phi}}]\,,
\label{dualGW}
\ee
which are known as the $\cN=2$, $d=3$ superspace action of the improved tensor
multiplet \cite{HKLR}. Note that analogous $\cN=1$, $d=4$ superspace action
of the improved tensor multiplet was constructed in \cite{LR}.
It is interesting to point out that (\ref{dualGW}) was obtained in \cite{KLL}
as a dual representation of the classical action of the Abelian
Gaiotto-Witten model \cite{GW}. Hence, the classical action of the Abelian
Gaiotto-Witten model in the representation (\ref{dualGW}) arises as the
leading term in the $\cN=4$ SYM effective action.

\subsection{$\cN=8$ SYM}
Consider the $\cN=8$ SYM model with the classical action
(\ref{N8}). The background-quantum splitting of the gauge superfield
(\ref{Vv}) is supplemented by the following splitting of the
(anti)chiral superfields,
\be
\Phi_i\to\Phi_i+g\phi_i\,,\qquad
\bar\Phi^i\to\bar\Phi^i+g\bar\phi^i\,,
\ee
with the corresponding `quantum' gauge transformations
\be
\delta \phi_i=i[\lambda,\frac1g\Phi_i+\phi_i]\,,\quad
\delta\bar\phi^i=i[\bar\lambda,\frac1g\bar\Phi^i+\bar\phi^i]\,,\quad
\delta\Phi_i=\delta\bar\Phi^i=0\,.
\label{q-tr2}
\ee

The gauge fixing functions are chosen in the form similar to
(\ref{f-f}),
\be
f= i\bar{\cal D}^2 v-\frac i2[\Phi_i,\bar{\cal
D}^2\square_-^{-1}\bar\phi^i]\,,\qquad
\bar f=i{\cal D}^2 v+\frac i2[\bar\Phi^i,{\cal D}^2\square_+^{-1}\phi_i]
\,.
\ee
When the background superfields are covariantly constant, $\bar{\cal
D}_\alpha\bar\Phi^i=0$, ${\cal D}_\alpha \Phi_i=0$, the quadratic
part of the action with respect to the quantum superfields takes
relatively simple form,
\bea
S_2+S_{\rm gf}&=&-\tr\int d^7z\left[
v(\square_{\rm v}+\bar\Phi^i\Phi_i)v+
\frac12\phi_i(\delta^i_j+\bar\Phi^i\Phi_j\square_-^{-1})\bar\phi^j
\right]\nn\\&&
+\frac14\left(
\tr\int d^5z\,\varepsilon^{ijk}\phi_i[\Phi_j,\phi_k] +c.c.
\right).
\eea
Here we denote $\bar\Phi^i\Phi_i v=[\bar\Phi^i[\Phi_i, v]]$. The
ghost superfield action is a simple generalization of
(\ref{gh-4}),
\be
S_{\rm gh}=\tr\int d^7z\left[
-b(1+\Phi_i\square_-^{-1}\bar\Phi^i)\bar c
+\bar b(1+\bar\Phi^i\square_+^{-1}\Phi_i)c
-\bar\varphi\varphi
\right].
\ee
As a result we see that the one-loop effective action is
relatively simple because it is defined by only one operator,
\be
\Gamma^{\cN=8}_{\rm SYM}=\frac i2{\rm Tr_v}\ln(\square_{\rm v}+\bar\Phi^i\Phi_i)
\,.
\label{N8EA}
\ee
The contributions from ghosts and chiral superfields cancel each
other at one loop for the covariantly constant background
similarly as it happens for the $\cN=4$, $d=4$ SYM theory.

For the gauge group SU$(N)$ spontaneously broken down to
U$(1)^{N-1}$ the gauge and chiral superfields are chosen
as in (\ref{V-back}) and (\ref{phi-back}). In this case the trace
of the logarithm in (\ref{N8EA}) is computed by standard methods
described in Sect.\ \ref{sect2.3},
\bea
\Gamma^{\cN=8}_{\rm SYM}&=& i\sum_{I<I}^N{\rm Tr_v}
\ln(\hat\square_{{\rm v}\,IJ}+\bar{\bf\Phi}_{IJ}^i{\bf\Phi}_{i\,IJ})
\label{G8}\\
&=&-\frac1\pi\sum_{I<J}^N\int d^7z
\int_0^\infty\frac{ds}{s\sqrt{i\pi s}}
\frac{{\bf W}_{IJ}^2\bar{\bf W}_{IJ}^2}{{\bf B}_{IJ}^3}e^{is({\bf G}_{IJ}^2
+\bar{\bf\Phi}_{IJ}^i{\bf\Phi}_{i\,IJ})}
\tanh\frac{s{\bf B}_{IJ}}{2}\sinh^2\frac{s{\bf B}_{IJ}}{2}\,.\nn
\eea

In the case when the gauge group SU$(N)$ is spontaneously broken
down to SU$(N-1)\times$U$(1)$, the background superfields should
be chosen as in (\ref{V-back1}) and (\ref{Phi-back2}). Then the
leading term in the effective action (\ref{G8}) is given by
\be
\Gamma^{\cN=8}_{\rm SYM}=\frac{3(N-1)}{32\pi}\int d^7z
\frac{{\bf W}^2\bar{\bf W}^2}{({\bf G}^2+\bar{\bf\Phi}^i{\bf\Phi}_i)^{5/2}}
+\ldots\sim \int d^3x\frac{(F^{mn}F_{mn})^2}{(f^{\ui} f_{\ui})^{5/2}}+\ldots\,,
\label{G8lead}
\ee
where $f^{\ui}$, $\ui=1,2,\ldots,7$ are the seven scalar fields in the
$\cN=8$ SYM theory and dots stand for the higher-order terms.
In \cite{DS} it was argued that the $F^4$ term in the $\cN=8$ SYM
effective action (\ref{G8lead}) is one-loop exact in the
perturbation theory, but it receives instanton corrections.

\subsection{$\cN=2$ Chern-Simons model}

Let us consider pure $\cN=2$ Chern-Simons theory with the
classical action (\ref{CS}). The background field method goes
along the same lines as in Sect.\ \ref{BFM}, but in eq.\ (\ref{Sf}) we
send $\alpha\to\infty$ and $\beta=1$ to
remove the term responsible for the gauge fixing in
the SYM part of the action. Then the quadratic part of the action
for the quantum gauge superfields reads
\be
S_2+S_{\rm gf}=\tr\int d^7z\, v H v\,,
\ee
with $H$ given in (\ref{H}).

The action for the Nielsen-Kallosh ghost vanishes because
$\varphi^2\equiv0$ for the anticommuting superfield. Hence, only
Faddeev-Popov ghosts contribute in the pure Chern-Simons theory.
As a result, the structure of one-loop effective action is given
by
\be
\Gamma_{\rm CS}^{\cN=2}=\Gamma_{H}+\Gamma_{\rm gh}\,,\qquad
\Gamma_{H}=\frac i2 \Tr_{\rm v} \ln H\,,\qquad
\Gamma_{\rm gh}=-i\Tr_+ \ln \square_+\,.
\ee
The operator $H$ is first order in space-time derivatives.
Therefore we need to square it,
\bea
\Gamma_H&=&\frac i4 \Tr_{\rm v}\ln H^2\,,\label{GHH}\\
H^2&=&-m^2\left[{\cal D}^m {\cal D}_m
+\frac i2(W^\alpha-\bar W^\alpha)({\cal D}_\alpha+\bar{\cal D}_\alpha)
\right],
\eea
where the mass $m^2=k^2g^4/(4\pi^2)$ was introduced in
(\ref{mass}).
Upon such a squaring we must care about the phase of the functional
determinant,
\be
\frac1{\sqrt{\det H}}=\frac1{|\sqrt{\det
H}|}\exp\left(\frac{i\pi}{2}\eta[V]\right),
\ee
where $\eta[V]$ is the so-called eta-invariant (see \cite{Witten} for
details in the non-supersymmetric case). It is a
background-dependent functional formally defined as
\be
\eta[V]=\frac12\lim_{s\to0} \sum_{i} \mbox{sign}
\lambda_i|\lambda_i|^{-s}\,,
\ee
where $\lambda_i$ are the eigenvalues of the operator $H$.
Fortunately, in \cite{BP} it was proved that
\be
\eta[V]=0\,.
\ee
Indeed, the non-vanishing value of the eta-invariant might lead only
to finite shifts of the Chern-Simons coupling constant $k$ because of
quantum divergences, but it is well known that there are no such
shifts in the Chern-Simons models with $\cN>1$ supersymmetry
\cite{KLL96}.

It is important to specify the background above which one
computes the quantum corrections. Recall that in the SYM theory we
used the constant field background constrained by (\ref{approx-2})
and (\ref{approx-3}). Such constraints provided us with a
consistent quantum field theory as such a background was a
solution of classical equations of motion. However, in the pure
$\cN=2$ Chern-Simons theory the equations of motion have only
trivial solutions with vanishing gauge superfield strengths.
Therefore, in quantizing the Chern-Simons theory we do not impose
any constraints on the background and compute the leading terms in
the derivative expansion of the effective action. In other words,
there is no Coulomb branch and we need to study the effective
action in the conformal branch when all fields are massless.

\subsubsection{Contributions from quantum vector superfields}
As the quantum vector superfields are massless, we need to
introduce an effective infrared cut-off $m$ to avoid IR
divergences. Then the effective action (\ref{GHH}) can be
represented as
\be
\Gamma_H[V]=\frac i4\Tr_{\rm v}\int_0^\infty \frac{ds}{s}e^{-m^2s}e^{-s H^2}.
\ee
Recall that $\Tr_{\rm v}{\cal O}$ for a differential operator $\cal O$
acting in the space of real superfields is computed by the rule
\be
\Tr_{\rm v}{\cal O}=\int d^7z\, {\cal O}\delta^7(z-z')|_{z=z'}\,,
\ee
where $\delta^7(z-z')=\delta^3(x-x')\delta^4(\theta-\theta')$.
Hence, to get non-vanishing result we have to accumulate exactly two derivatives ${\cal
D}_\alpha$ and two $\bar{\cal D}_\alpha$ ones on the
delta-function from the expansion of $e^{-sH^2}$. Such a decomposition
is straightforward. The expression with the minimal number of
superfield strengths reads
\be
\Gamma_H=-\frac{1}{256\pi m^5}\int d^7z
\left({\cal W}^{a \alpha} {\cal W}^{b}_\alpha {\cal W}^{c\beta}{\cal W}^{d}_\beta
 -
\frac12{\cal W}^{a\alpha} {\cal W}^{b\beta}{\cal W}^{c}_\alpha
{\cal W}^{d}_\beta\right){\bf f}_{abcd}+O(m^{-6})\,,
\ee
where ${\cal W}^{a\alpha}\equiv
(W^{a\alpha}-\bar{W}^{a\alpha})T_a$, $[T_a, T_b]=f_{abc}T_c$, and
\be
{\bf f}_{a_1 a_2 a_3 a_4}=f_{b_1 a_1 b_2}f_{b_2 a_2
b_3}f_{b_3 a_3 b_4} f_{b_4a_4b_1}\,.
\ee
Note that these terms do not have Abelian analogs.
They simply vanish in the Abelian case.

\subsubsection{Contributions from ghost superfields}
Consider one-loop contributions from the Faddeev-Popov
ghost superfields,
\be
\Gamma_{\rm gh}=-i\Tr_+\ln\square_+=-i\int d^5z\int_0^\infty \frac{ds}{s}
e^{-sm^2}K_+(z|s)\,,
\label{461}
\ee
where $K_+(z|s)$ is the heat kernel for the chiral box operator
(\ref{square+}) with coincident superspace points,
\be
K_+(z|s)=\tr e^{-s\square_+}\delta_+(z,z')|_{z=z'}{\bf
1}\,.
\ee
Note that for the constant superfield strengths this heat kernel
has exact expression (\ref{K+}). In the present section we need
the value of this heat kernel beyond the constant field
approximation. In this case only the lower order terms in the
series expansion over the parameter $s$ can be found exactly.

Using the relation (\ref{square+}) we can restore the full
superspace measure in (\ref{461}), but for the derivative of the
heat kernel,
\be
\int d^5z \frac d{ds} K_+(z|s)=\frac14\tr \int d^7z\, {\cal
D}^2 e^{-s\square_+}\delta_+(z,z')|_{z=z'}\,.
\ee
Let us make a series decomposition
\be
\tr{\cal D}^2 e^{-s\square_+}\delta_+(z,z')|_{z=z'}
=\frac1{(4\pi s)^{3/2}}\sum s^n C_n(z)\,,
\label{denote}
\ee
with coefficients $C_n(z)$ being superfields in full superspace.
Then we have
\bea
\Gamma_{\rm gh}&=&-\frac i4\int d^7z\int_0^\infty ds\, e^{-m^2 s}
\sum\frac{s^n C_n(z)}{(n-\frac12)(4\pi s)^{3/2}}\nn\\
&=&-\frac i{32\pi^{3/2}} \sum_{n=1}^\infty \frac{\Gamma(n-1/2)}{(n-1/2)m^{2n-1}}
\int d^7z\, C_n(z)\,.
\eea
As a result, we need to compute the superfield coefficients
$C_n(z)$ in the series decomposition of the lhs in (\ref{denote}).
These coefficients play the role of the superfield Schwinger-DeWitt
coefficients \cite{BK-book}.

To compute the coefficients $C_n$ in (\ref{denote}) we use the
Fourier representation for the chiral delta-function,
\be
\delta_+(z,z')=-4\int \frac{d^3p}{(2\pi)^3}d^2\eta \,
e^{ip^m \zeta_m+\eta^\alpha\zeta_\alpha}\,,
\ee
where $\zeta^m$ and $\zeta^\alpha$ are the components of
supersymmetric interval (\ref{interval}). The coefficient $-4$ here
is due to our normalization of the integration measure, $\int d^2\eta\, \eta^2=1$.
Now we act by the
operators ${\cal D}^2$ and $\square_+$ on $e^{ip^m
\zeta_m+\eta^\alpha\zeta_\alpha}$ and consider the limit of
coincident superspace points,
\be
\tr{\cal D}^2 e^{-s\square_+}\delta_+(z,z')|_{z=z'}
=-4\tr\int\frac{d^3p}{(2\pi)^3}d^2\eta\, X^\alpha X_\alpha
e^{-s(X^m X_m+iX^\alpha W_\alpha+G^2)}|_{z=z'}\,,
\label{K}
\ee
where
\be
X_m={\cal D}_m+ip_m\,,\qquad
X_\alpha={\cal D}_\alpha +\eta_\alpha -p_{\alpha\beta}\bar
\zeta^\beta\,.
\ee
Note that the last term $p_{\alpha\beta}\bar\zeta^\beta$ in
$X_\alpha$ does not contribute to the expansion of the heat
kernel as it vanishes in the limit of coincident superspace
points. Now we expand the exponent in a series and integrate over
the momenta, e.g.,
\be
\int\frac{d^3p}{(2\pi)^3}e^{sp^2}=\frac{i}{(4\pi s)^{3/2}}\,.
\ee

It is clear that $C_0$ is independent of the background and the
decomposition starts with $C_1$. The latter appears from
$-s(\square+G^2)$ with the factor $i/(4\pi s)^{\frac32}$ which is
common in the expansion. In the next order of expansion after
integration over $p$ we have exactly $+s\square$ that cancels gauge
non-invariant contribution and then we find \be \label{e1} C_1=4i\tr
G^2\,. \ee As a result, the leading contribution to the effective
action due to ghost superfields has the form of Yang-Mills action in
the $\cN=2$, $d=3$ superspace,
\be
\Gamma_{\rm gh}=\frac{1}{8\pi^2
m}\tr\int d^7z\,G^2 +O(m^{-2})=-\frac1{16\pi^2 m}\tr\int d^5z\,
W^\alpha W_\alpha+O(m^{-2}) \,.
\ee
This demonstrates that the Yang-Mills term is generated in the
effective action of pure $\cN=2$ supersymmetric Chern-Simons
theory by the ghost superfield loop.
The appearance of this term in the effective action is not
surprising as we break the conformal invariance and topological
nature of pure $\cN=2$ Chern-Simons theory. Clearly, this term
vanishes on shell for $W_\alpha=0$. Similar $F^2$ terms in the
off-shell effective action of non-supersymmetric Chern-Simons theory were
discussed in \cite{CSW,CSW1,CS-review}.

The procedure developed in this section allows one to compute all other superfield
coefficients in (\ref{denote}) which contain higher orders of
superfield strengths, but this task is more tedious for $C_n$ with
$n>1$.

\section{Conclusions}
We have reviewed the construction of the background field method for
gauge field theories in the $\cN=2$, $d=3$ superspace and
demonstrated its power for calculating the low-energy effective
action for $\cN=2,4,8$, $d=3$ super Yang-Mills models and
$\cN=2$ super Chern-Simons model. The background-field-depended
operators of quadratic fluctuations, which represent the key elements of
the background field formalism, are exactly found in the $\cN=2$ super
Yang-Mills and Chern-Simons models for arbitrary gauge superfield
background. The structure of one-loop effective action for these
models is discussed in details. We have developed the $\cN=2$, $d=3$
superfield heat kernel technique and applied it for calculating
the low-energy effective actions in a form preserving manifest gauge
invariance and $\cN=2$ supersymmetry.

For constant gauge superfield background the heat kernel for the
operator of quadratic fluctuations in the SYM theory was exactly
found. This allows us to find the one-loop effective action in the
$\cN=2$, $\cN=4$ and $\cN=8$ SYM for such a background. However, in
the pure $\cN=2$ Chern-Simons theory only vanishing gauge superfield
background is allowed as a solution of classical equations of
motion. Therefore we compute the effective action in the $\cN=2$
Chern-Simons theory only in the conformal branch when all the gauge
degrees of freedom are massless. In this case we consider the
arbitrary background superfield and  compute the leading terms in
the effective action of this model containing lowest number of gauge
superfields. We show that such off-shell effective action contains
the Yang-Mills term which appears due to ghost superfield
contributions.

The methods developed in \cite{BPS1,BPS2,BP} and reviewed in this
paper can be applied to a wide class of three-dimensional extended
supersymmetric gauge theories. For example, it would be interesting
to study the higher loop low-energy effective action in
three-dimensional $\cN=2$ and $\cN=4$ SYM theories with matter
and in the models containing both SYM and Chern-Simons terms
together. More importantly, it is tempting to study the low-energy
effective action of ABJM-like models which could correspond to the
effective action of the M2 brane on the AdS$_4\times$S$^7$
background. The latter can provide one more non-trivial evidence of
the AdS$_4$/CFT$_3$ correspondence.

\vspace{30pt}
\noindent
{\bf Acknowledgments}\\[3mm]
The authors are grateful to the RFBR grant Nr.\ 12-02-00121 and LRSS
grant Nr.\ 224.2012.2 for partial support. I.L.B.\ and I.B.S.\
acknowledge the support from the RFBR-Ukraine grant Nr.\ 11-02-90445
and DFG grant LE 838/12/1. The work of I.B.S.\ was also supported by
the Marie Curie research fellowship Nr.\ 236231,
``QuantumSupersymmetry''. N.G.P.\ acknowledges the support from RFBR
grant Nr.\ 11-02-00242.



\end{document}